\documentclass[aps,prb,twocolumn,showpacs,superscriptaddress]{revtex4}
\usepackage{graphicx}
\usepackage{color}
\renewcommand{\vec}[1]{\mathbf{#1}}
\begin{document}
\title{Ferromagnetic phase in the polarized two-species bosonic Hubbard Model}

\author{Kalani Hettiarachchilage} 
\affiliation{Department of Physics and Astronomy, Louisiana State University, Baton Rouge, Louisiana 70803, USA}
\affiliation{Center for Computation and Technology, Louisiana State University, Baton Rouge, LA 70803, USA}
\author{Val\'ery G.~Rousseau}
\affiliation{Department of Physics and Astronomy, Louisiana State University, Baton Rouge, Louisiana 70803, USA}
\affiliation{Center for Computation and Technology, Louisiana State University, Baton Rouge, LA 70803, USA}
\author{Ka-Ming Tam}
\affiliation{Department of Physics and Astronomy, Louisiana State University, Baton Rouge, Louisiana 70803, USA}
\affiliation{Center for Computation and Technology, Louisiana State University, Baton Rouge, LA 70803, USA}
\author{Mark Jarrell}
\affiliation{Department of Physics and Astronomy, Louisiana State University, Baton Rouge, Louisiana 70803, USA}
\affiliation{Center for Computation and Technology, Louisiana State University, Baton Rouge, LA 70803, USA}
\author{Juana Moreno}
\affiliation{Department of Physics and Astronomy, Louisiana State University, Baton Rouge, Louisiana 70803, USA}
\affiliation{Center for Computation and Technology, Louisiana State University, Baton Rouge, LA 70803, USA}

\date{\today}

\begin{abstract}
We recently studied a doped two-dimensional bosonic Hubbard model with two hard-core species, 
with different masses, using quantum Monte Carlo simulations [Phys. Rev. B 88, 161101(R) (2013)]. 
Upon doping away from half-filling, we find several distinct phases, including a 
phase-separated ferromagnet with Mott behavior for the heavy species and 
both Mott insulating and superfluid behaviors for the light species. Introducing polarization, an imbalance 
in the population between species, we find a fully phase-separated ferromagnet.
This phase exists for a broad range of temperatures and polarizations. By using finite 
size scaling of the susceptibility, we find a critical exponent which is consistent with the two-dimensional Ising 
universality class. Significantly, since the global entropy of this phase is 
higher than that of the ferromagnetic phase with single species, its experimental observation in cold atoms may be feasible.

\end{abstract}

\pacs{02.70.Uu,05.30.Jp}
\maketitle
\section{Introduction}
One of the frontiers of condensed matter physics is the study of competing 
quantum phases such as coexistent and inhomogeneous phases, quantum criticality, 
and secondary ordered phases close to quantum critical points.~\cite{Yunoki,Yunoki2,Uehara,Elbio1,Coleman,Si} 
These exotic phenomena in strongly correlated 
systems occur due to the competition and cooperation between the spin, charge, lattice, 
and orbital degrees of freedom.~\cite{Elbio} Unfortunately, it is often difficult 
to differentiate the effect of these degrees of freedom in real materials. However, 
the advance of optical lattice experiments provides a tantalizing opportunity 
to study competing phases via controlled external parameters.~\cite{Jaksch,Hofstetter,Esslinger}

The experimental tunability of Hamiltonian parameters using 
laser and magnetic fields~\cite{Timmermans99, kohler:1311} allows the realization of 
strongly correlated model Hamiltonians. The realization of 
the Bose-Hubbard model using ultra-cold atoms on optical lattices~\cite{Greiner} 
has led to the observation~\cite{Fisher,Batrouni} of the Mott-insulator to superfluid phase transition. The Mott insulator 
phase is characterized by commensurate occupations, gapped excitations and incompressibility in the strong 
coupling regime. The superfluid phase is characterized by Bose-Einstein condensation, gapless 
excitations and finite compressibility in the weak coupling regime. 

This success has spurred interest in mixtures of atoms which can give rise to even more interesting 
and complex phases.  These include mixtures of bosonic and fermionic atoms~\cite{Modugno,Albus,Modugno1,Ospelkaus} 
(a Bose-Fermi mixture) and mixtures of two different bosonic species (a Bose-Bose mixture).~\cite{Roati,Thalhammer,Papp}
Moreover, experimental studies of $^{85}Rb$-$^{87}Rb$, $^{87}Rb$-$^{41}K$, $^{6}Li$-$^{40}K$ and different alkaline 
earth mixtures in optical lattices~\cite{Catani,Taie,Taglieber} have motivated theoretical studies of the two species
Bose-Hubbard model.~\cite{Altman,Soyler,Soyler2,Stephen,Andrii,Lv,Trousselet,Kuno,Kalani} The 
zero-temperature phase diagram of the two-dimensional, two-species, hard-core bosonic Hubbard model has been studied 
at half-filling using a combination of mean field and variational methods,~\cite{Altman} and by means of quantum
Monte Carlo simulations.~\cite{Soyler} The rich phase diagram found at half-filling in these studies shows ordered
Mott insulating phases including anti-ferromagnetic and super-counter-fluid phases in the strong interaction limit. On the other 
hand, superfluid and antiferromagnetic/superfluid phases are found in the weak interaction limit.~\cite{Altman,Soyler,Soyler2,Stephen,Andrii} 
Recently, we have included doping dependence as a control parameter to study this model using quantum Monte Carlo  simulations. 
We found several distinct phases including a normal liquid at higher temperatures, an antiferromagnetically 
ordered Mott insulator, and a region of coexistent antiferromagnetic and superfluid order near half-filling.~\cite{Kalani}
We also reported a small dome containing a phase-separated
ferromagnetic phase away from half-filling at zero polarization.

Though the realization of quantum magnetic phases has gained
significant attention, the prominent experimental challenge is to reach the low 
temperatures and entropies needed to observe these phases. Several different experimental techniques 
have been proposed to reach such low entropies.~\cite{Monroe,Popp,Li} Interestingly, a Bose-Fermi 
mixture may be used to squeeze the entropy of a Fermi gas into the surrounding Bose gas.~\cite{Ho-Zhou} 
This can leave a low entropy heavy Fermi gas by evaporating the entropy absorbed by the light Bose gas. 
The relatively high global entropy of the phase-separated
ferromagnetic phase we found away from half-filling in the two species Hubbard model~\cite{Kalani} 
suggests that this ferromagnet should be easier to access experimentally. 

In the experimental setup of boson-boson mixtures, the two species are not always 
perfectly balanced.~\cite{Thalhammer,Plich} The evaporative cooling leads to net losses of one of the species, due
to the difference in the effective depth of the traps. This can be adjusted by loading different
number of atoms for different species into the trap.~\cite{Plich} This procedure can also be used to set an imbalance
amount of atoms for the two species. Most of the previous theoretical or numerical
studies on the two-species Bose-Hubbard model do not directly address this imbalance
in the experimental conditions.

In this paper, we explore the extent of the phase-separated ferromagnetic phase as a function 
of {\it finite} polarization, i.e., with a different population for each species. When the polarization 
is positive (more of the light than heavy particles) we find a larger region of the ferromagnetic 
phase-separated order, with higher transition temperatures and greater extent in doping.
Since this ferromagnetic phase exists for a broad range of sufficiently high temperatures and polarizations 
together with high global entropies, experimental observation in cold atoms may be achievable.

This manuscript is organized as follows. In section \ref{Sec:Model} we describe our model and 
method. The density versus polarization phase diagram at low temperature is studied in 
section \ref{Sec:DPphasediagram}.  In section \ref{Sec:PTphasediagram} we discuss the temperature 
versus polarization phase diagram along an optimal superfluid or maximum ferromagnetic phase line. 
The momentum distribution of the ferromagnetic and superfluid phases are presented in 
section \ref{Sec:Momentum}. In section \ref{Sec:Entropy} we calculate the entropy of the ferromagnetic, 
antiferromagnetic and superfluid phases. Finally we conclude in section \ref{Sec:Conclusion}.

\section{Model and method} \label{Sec:Model}
The Hamiltonian for the two-species Hubbard model with hard-core heavy, {\it a}, and light, {\it b}, 
bosons confined on a two-dimensional square lattice takes the form:
\begin{eqnarray}
  \label{Eq:Hamiltonian} \nonumber \hat\mathcal H &=& -t_{a}\sum_{\langle i,
j\rangle}\Big(a_i^\dagger a_j^{\phantom\dagger}+H.c.\Big)\\
& & -t_{b}\sum_{\langle i,j\rangle}\Big(b_i^\dagger b_j^{\phantom\dagger}
+H.c.\Big)+{U_{ab}\sum_i \hat n_i^{a} \hat n_i^{b}},
\end{eqnarray}
where $a_i^\dagger$ ($b_i^\dagger$) and $a_i$ ($b_i$) are the creation and annihilation 
operators of hard-core bosons {\it a} and ({\it b}), respectively, with number operators 
$\hat n_i^{a}=a_i^\dagger a_i^{\phantom\dagger}$, $\hat n_i^{b}=b_i^\dagger b_i^{\phantom\dagger}$.  
The sum $\sum_{\langle i,j\rangle}$ runs over all distinct pairs of first neighboring 
sites $i$ and $j$, $t_a (t_b)$  is the hopping integral between sites $i$ and $j$ for 
species {\it a} ({\it b}), and $U_{ab}$ is the strength of the on-site interspecies repulsion. 

We perform a quantum Monte Carlo study of the model (\ref{Eq:Hamiltonian}) by using the 
Stochastic Green Function algorithm~\cite{SGF,DirectedSGF} with global space-time 
updates~\cite{SpaceTime} to solve the canonical ensemble on $L \times L$ lattices. 
We focus on the polarized phase diagram, with polarization $P=\frac {N_b -N_a}{L^2}$ and 
total density $\rho=\frac {N_b +N_a}{L^2}$, with $N_a$ and $N_b$ the number of heavy 
{\it a} and light {\it b} particles, respectively. For the other parameters we use the 
same values as in Ref.~\onlinecite{Kalani}, namely $t_a=0.08\, t$, $t_b=t$, and $U_{ab}=6t$, 
where $t=1$.

\section{Phase diagram at low temperature} \label{Sec:DPphasediagram}
\begin{figure}[!htbp] 
 \centerline{\includegraphics[width=0.5\textwidth]{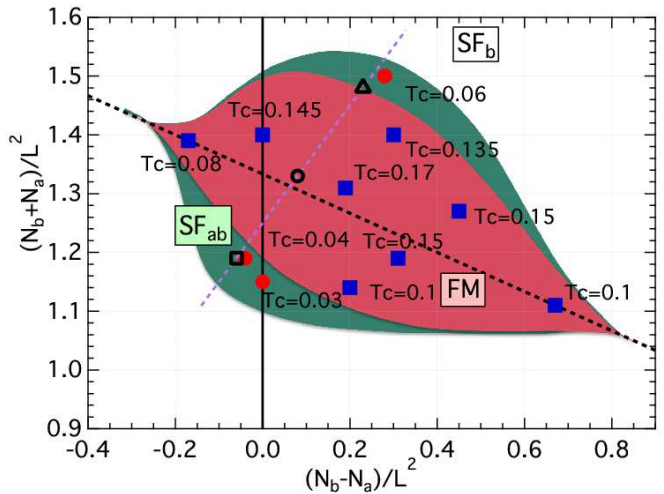}} 
 \caption
    {(Color online) 
The total density, $\rho=\frac{N_b+N_a}{L^2}$, versus polarization, $P=\frac{N_b-N_a}{L^2}$, phase 
diagram at very low temperature. The transition temperatures 
associated with the data points are obtained from finite-size scaling calculations.
The boundaries between the phases are estimated for $\beta t=60$ and $L=10$ with 
$t_a=0.08t, t_b=1.00t$, and $U_{ab}=6t$. The boundaries may slightly change for the ground state. 
The red area shows the phase-separated ferromagnet (FM). The green area shows the region
of superfluidity of both {\it a} and {\it b} species (SF$_{ab}$). The white region represents the superfluidity 
of light {\it b} particles (SF$_b$)except that the system is an antiferromagnet at half-filling ($\rho=1$ 
and $P=0$), and there is an antiferromagnetic to superfluid phase-separated 
region near half-filling ($1.0<\rho<1.1$ and $P=0$) as discussed in Ref.~\onlinecite{Kalani}. Both particles 
are in a Mott insulating phase whenever their individual densities are integers ($0$ or $1$).
The blue squares indicate the transition temperatures from the light species superfluid to the ferromagnetic phase.
The red circles correspond to the transition temperatures from the light species superfluid to the phase where both species
are superfluid. The black dotted line,
$N_b+\frac{N_a}{2}=L^2$, follows the highest ferromagnetic critical temperature. The phase diagram 
as a function of temperature along this black dotted line is shown in Fig.~\ref{Fig:DensityTemp}. The 
momentum distributions shown in Fig.~\ref{Fig:Momentum} are calculated along the purple dotted line ($N_a=0.625L^2$), which
intersects the black dotted line. The momentum distributions for heavy and light species for 
three points along the purple dotted line are shown in the bottom panels of Fig.~\ref{Fig:Momentum}. 
The black line is the zero polarization axis. 
}
\label{Fig:DensityDensity}
\end{figure}

Figure~\ref{Fig:DensityDensity} displays the total density $\rho$ versus polarization $P$ phase diagram at
low temperature. In the thermodynamic limit, a ferromagnetic phase exists in a broad region of
densities (red area), heavy {\it a} and light {\it b} particle superfluidity exists in a smaller region of densities (green area), 
and superfluidity of light {\it b} particles (with heavy {\it a} particles in the normal state) appears in most of 
the rest of the phase diagram (white area). Along the zero polarization axis there is an antiferromagnetic phase at 
half-filling ($\rho=1$ and $P=0$), and an antiferromagnetic to superfluid phase-separated 
region for $1.0<\rho<1.1$ and $P=0$ as discussed in Ref.~\onlinecite{Kalani}.  
The black dotted line, $N_b+\frac{N_a}{2}=L^2$, follows the highest ferromagnetic critical temperatures (optimal superfluid 
line). Along this line the system shows fully phase-separated regions of average local densities $n_a\sim 0$ together with
$n_b\sim 1$, and $n_a\sim1$ with $n_b\sim 0.5$.  Therefore, the number of light particles, $N_b$, is given as 
$N_b= (L^2 -N_a) + \frac{N_a}{2}$ (or $\frac{N_a}{2}+N_b=L^2$). In our previous study we did not distinguish between
the phase where only the light particles are superfluid from the one where both species are superfluid. The ferromagnetic 
phase boundaries at zero polarization have also changed slightly.

The blue squares in Fig.~\ref{Fig:DensityDensity} indicate the transition temperature from the light species superfluid to 
the ferromagnetic phase for the given densities and polarizations. To find these transition 
temperatures we calculate the ferromagnetic susceptibility for different system sizes and perform a finite-size scaling.
The susceptibility is given as $\mathcal \chi(\vec k) = \big\langle \vert \mathcal A(\vec k)\vert^2\big\rangle -\vert\big\langle \mathcal A(\vec k)\big\rangle\vert^2$ with
\begin{eqnarray}
  \label{Eq:Susceptibility} \mathcal A(\vec k) &=& \frac{1}{\beta}\int_0^{\beta} \sum_j e^{i\vec k\cdot \vec r_j}(n_j^a(\tau)-n_j^b(\tau))\,d\tau.
\end{eqnarray}
Following Ref.~\onlinecite{Banos-etal-2012} we calculate the ferromagnetic susceptibility ratio, $\mathcal R$, defined as
\begin{eqnarray}
  \label{Eq:SusceptibilityRato}  \mathcal R&=& \frac{\chi(0,\varepsilon)+\chi(0,-\varepsilon)+\chi(\varepsilon,0)+\chi(-\varepsilon,0)}{\chi(\varepsilon,\varepsilon)+\chi(-\varepsilon,\varepsilon)+\chi(\varepsilon,-\varepsilon)+\chi(-\varepsilon,-\varepsilon)}, 
\end{eqnarray}
where $\varepsilon=\frac{2\pi}{L}$. We impose all the point group symmetries in $\vec k$-space near $\vec k\sim0$ to both the
numerator and denominator to reduce the statistical noise associated with quantum Monte Carlo sampling.

\begin{figure}[!htbp] 
 \centerline{\includegraphics[width=0.5\textwidth]{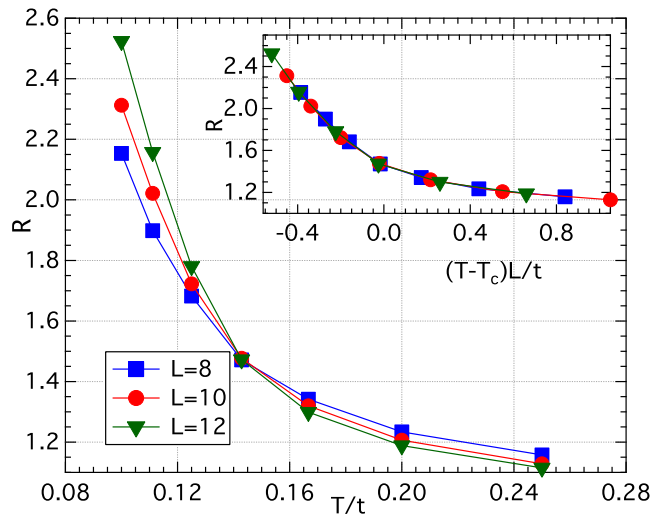}} 
 \caption
    {(Color online)
Scaling behavior of the ferromagnetic susceptibility for the continuous transition from light species superfluid
to ferromagnet at $\rho=1.4$ and $P=0$.  The susceptibility ratios, $\mathcal R$ (Eq.~\ref{Eq:SusceptibilityRato}),
versus temperature, $T/t$, for different system sizes cross at the critical temperature, $T_c=0.145t$. The inset shows 
the scaling near the critical temperature. The curves collapse onto a single curve with the critical exponent of correlation 
length $\nu=1$. The data points are based on simulation results, the lines are guides to the eye.        
     }
\label{Fig:Ferro}
\end{figure}

The scaling behavior of the ferromagnetic transition temperature is shown in Fig.~\ref{Fig:Ferro} where the susceptibility 
ratio $\mathcal R$ is plotted for different system sizes as a function of temperature $T/t$ at $\rho=1.4$ and $P=0$.
In Fisher scaling~\cite{Fisher1,Fisher2,Fisher3}, the susceptibility at small wavenumber should scale as 
$\chi \sim L^{\frac{\gamma}{\nu}}g(L^{\frac{1}{\nu}}(T-T_c))$, where $\gamma$ and $\nu$ are 
the critical exponents for the ferromagnetic susceptibility and correlation length, respectively. 
By looking at the ratio of the susceptibilities $\mathcal R$, the $L^{\frac{\gamma}{\nu}}$
factor is canceled. At the transition, the scaling function $g(0)$ is independent of $L$. 
Thus, the susceptibility ratio $\mathcal R$ versus temperature $T$ for different system
sizes should cross at the critical temperature $(T=T_c=0.145t)$ as it is shown in Fig.~\ref{Fig:Ferro}. 
By choosing the critical exponent of the correlation length as $\nu=1$, the
value for a two-dimensional Ising transition,  we find that the curves collapse onto one curve 
near the critical temperature (c.f., the inset of Fig.~\ref{Fig:Ferro}). 
If the transition is second order by considering  symmetry arguments, it should belong to the Ising 
universality class. However, if we understand the polarized model as a Ising system within an external magnetic 
field, it is possible that the ferromagnetic  transition is first order. Since it is very difficult to distinguish 
between first and second order phase transitions with our finite size calculations, we can not clarify this issue.

\begin{figure}[!htbp] 
 \centerline{\includegraphics[width=0.5\textwidth]{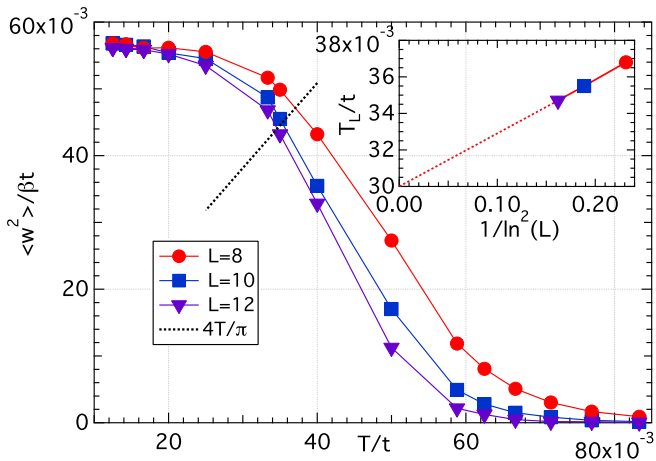}} 
 \caption
    {(Color online)
Winding number of heavy {\it a} particles as a function of temperature for different system 
sizes at $\rho=1.14$ and $P=0$.  The black dotted line corresponds to $\frac{4T}{\pi}$ 
and is used to find the crossing temperature for different system sizes. 
The inset shows the finite size scaling of the crossing temperatures to find the superfluid 
critical temperature, $T_c=0.03t$, in the thermodynamic limit. The data points are based 
on simulation results, the lines are guides to the eye.        
     }
\label{Fig:SFA}
\end{figure}

The red circles in Fig.~\ref{Fig:DensityDensity} indicate the superfluid transition temperature 
for the heavy species. The scaling behavior of the superfluid to normal liquid
transition should follow that of the Kosterlitz-Thouless continuous transition.  We note that the 
Hamiltonian (\ref{Eq:Hamiltonian}) satisfies the condition (28) of Ref.~\onlinecite{Superfluid},
which allows one to relate the superfluid density to the fluctuations of the winding 
number.~\cite{Pollock} 
In Fig.~\ref{Fig:SFA}, we show the winding number of the {\it a} 
particles, $\langle W^2 \rangle$, as a function of temperature, $T/t$, for different 
system sizes. The order parameter, the superfluid density, has a universal jump of 
$\langle W^2 \rangle=\frac{4}{\pi}$ at the critical point.~\cite{Nelson} The black dotted 
line shows  $\frac{4}{\pi}T$ as a function of temperature. We read the crossing 
temperature, $T_L$, for different system sizes. Then we use the relation between 
the crossing temperature, $T_L$, and the cluster size, 
$L$,  $T_L-T_c(\infty)\propto \displaystyle \frac {1}{ln^{2}(L)}$,~\cite{Boninsegni05} 
to find the critical temperature, $T_c$, in the thermodynamic limit. The inset of
Fig.~\ref{Fig:SFA} displays this scaling. We find $T_c=0.03t$ at $\rho=1.14$ and $P=0$. 

\section{Phase diagram on the optimal superfluid line} \label{Sec:PTphasediagram}
\begin{figure}[!htbp] 
 \centerline{\includegraphics[width=0.5\textwidth]{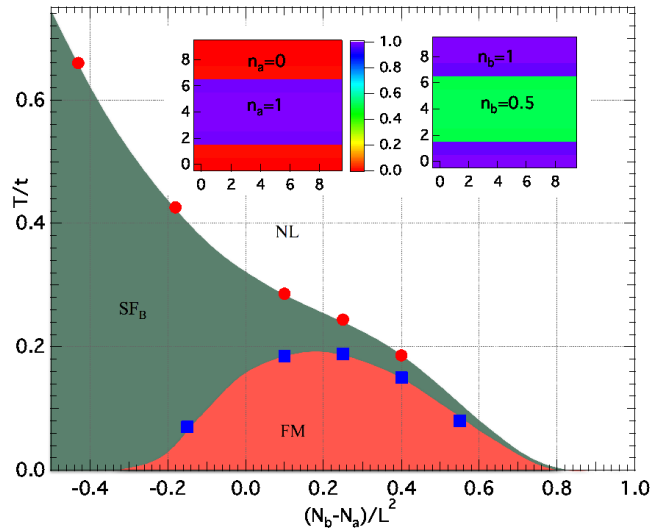}}
  \caption
    {(Color online) 
The temperature, $T/t$, versus polarization, $P=\frac{N_b-N_a}{L^2}$, phase diagram when $N_b+\frac{N_a}{2}=L^2$. 
The abscissa extends from $\rho=1, P=1$ ($N_a=0$, $N_b=L^2$) to $\rho=1.5, P=-0.5$ ($N_a=L^2$, $N_b=L^2/2$). 
The orange area corresponds to the phase-separated ferromagnet (FM). The green area is the region where the light 
{\it b} species displays  superfluidity (SF).   The white area is the normal liquid (NL). The 
blue squares and red circles indicating the boundaries between the phases are 
calculated by finite size scaling, see section \ref{Sec:DPphasediagram}. The curves are guides to the eye. 
Since it is difficult to perform finite size scaling at very low temperatures, the edges of the 
phase diagram are estimations of the transition temperatures based on results of small clusters.
The inset shows a snapshot of the average local densities versus lattice coordinates for $L=10$, $\rho=1.25$, 
$P=0.25$ and $\beta t=80$.  Left panel: For {\it a} particles, the red regions have  $\langle n_i^a\rangle \sim 0$ 
while the occupation of the blue region is $\langle n_i^a\rangle \sim 1$. Right panel: For {\it b} particles, the
blue regions have  $\langle n_i^b\rangle \sim 1$ while the occupation of the green region is $\langle n_i^b\rangle \sim 0.5$. 
The ferromagnetic phase separation occurs when the heavy species is in a Mott insulating state while the light 
one displays regions with either Mott insulating or superfluid behaviors.
    }
\label{Fig:DensityTemp}
\end{figure}

To better understand the phases of the polarized model we investigate snapshots 
of the average local densities. From the snapshots we propose that superfluid 
and ferromagnetic states are optimal along the black dotted line shown in 
Fig.~\ref{Fig:DensityDensity}, where $\frac{N_a}{2}+N_b=L^2$, with $L^2$ the lattice 
size, and $N_a$ and $N_b$ the number of heavy and light atoms, respectively. 
Along this line the system shows fully phase-separated regions with average local densities $n_a\sim 0$ and 
$n_b\sim1$ in the Mott region, and $n_a\sim1$ and $n_b\sim 0.5$ in the Mott/superfluid region.
The inset of Fig.~\ref{Fig:DensityTemp} displays snapshots of these average local 
densities for heavy (left panel) and light (right panel) particles. Physically, this 
optimal line is driven by the fact that a superfluid with $n_b\sim 0.5$ gains the most 
energy per particle.  As an example, for $N_a=50$, $N_b=75$, $L=10$
($\rho=1.25$ and $P=0.25$), half of the lattice is filled with {\it a} particles and 
the other half with {\it b} particles. The 25 remaining {\it b} 
particles will occupy the region filled by {\it a} particles and $n_b=\frac{25}{50}=0.5$ 
in that region. This reasoning is valid along this optimal superfluid line. However, 
when the system deviates far from $N_a=50\%$ of the number of lattice sites, it is difficult 
to stabilize small and large phase-separated regions. In this case, the pattern may 
break. This also explains why the ferromagnetic phase-separated phase is more stable 
for positive polarizations around $\rho =1.25$ and $P=0.25$. At half-filling of the 
heavy particles, this pattern is more stable since there are two large phase-separated
regions reducing surface effects. 

\begin{figure}[!htbp] 
 \centerline{\includegraphics[width=0.5\textwidth]{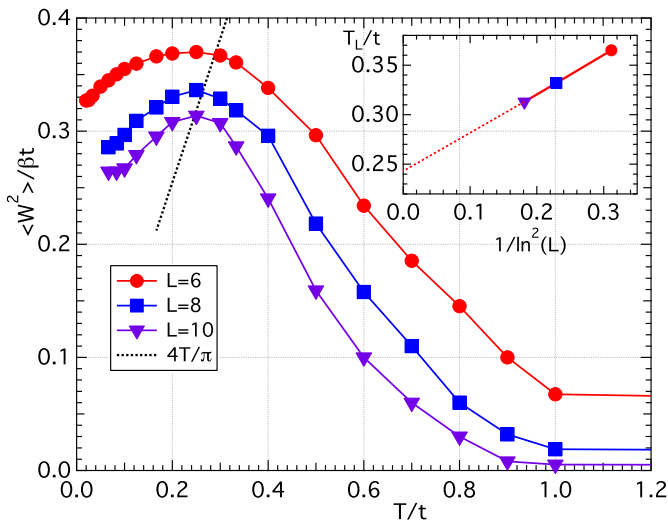}}
  \caption
    {(Color online) 
Winding number of light {\it b} particles as a function of temperature for different system sizes 
at $\rho=1.25$ and $P=0.25$. The black dotted line shows $\frac{4T}{\pi}$ and it is  used to 
find the crossing temperatures for different system sizes. The lowering of the superfluid density at low 
temperatures occurs when the system enters the ferromagnetic phase where 
the light species displays both superfluid and Mott behaviors. The inset shows 
the finite size scaling of the crossing temperatures to find the superfluid critical temperature, 
$T_c=0.245t$, in the thermodynamic limit for the continuous transition. The data points are 
based on simulation results, the lines are guides to the eye.      
    }
\label{Fig:SFB}
\end{figure}

Fig.~\ref{Fig:DensityTemp} shows the temperature, $T/t$, versus polarization, $P=\frac{N_b-N_a}{L^2}$, phase 
diagram on the optimal superfluid line, $\frac{N_a}{2}+N_b=L^2$. The blue squares are the ferromagnetic 
transition temperatures found by scaling as discussed in section \ref{Sec:DPphasediagram}. 
The red circles indicate the transition 
temperatures for light species superfluid. Again, the scaling behavior of this light particle
superfluid to normal liquid transition follows that of the Kosterlitz-Thouless transition 
as discuss earlier for the  heavy particles.  In  Fig.~\ref{Fig:SFB}, 
we show the winding of the {\it b} particles, $\langle W^2 \rangle$, as a function of temperature, $T/t$, 
for different system sizes. 
We find $T_c=0.245t$ at $\rho=1.25$ and $P=0.25$ as shown in Fig~\ref{Fig:SFB}. 

\begin{figure}[!htbp] 
 \centerline{\includegraphics[width=0.5\textwidth]{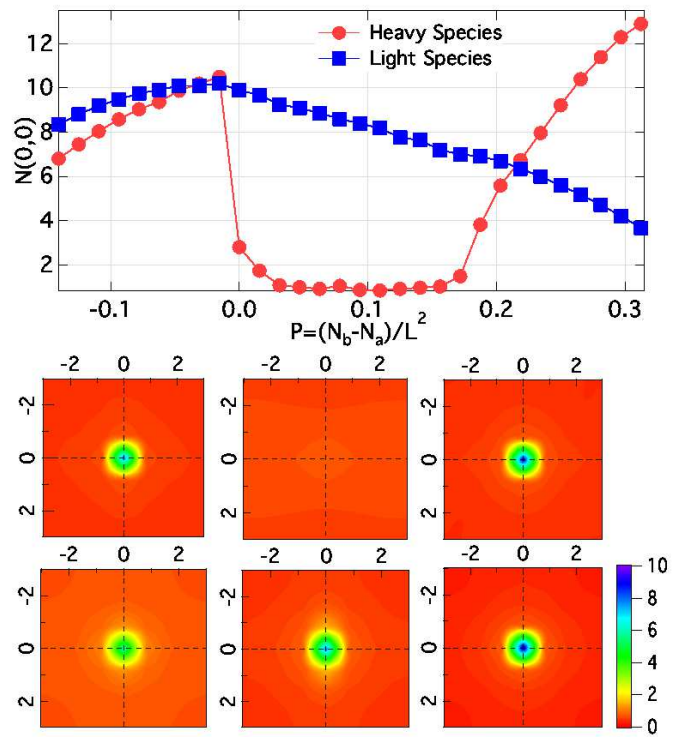}} 
 \caption
    {(Color online) 
The $\vec k$-space momentum distribution for $\beta t=50$ and $L=8$. Top panel: The momentum distribution at 
zero momentum $\vec k = (0,0)$ as a function of polarization for heavy (red circles) and 
light (blue squares) species along the purple dotted line shown in Fig.~\ref{Fig:DensityDensity}. 
The data points are based on simulation results, the lines are guides to the eye. Bottom panels: 
The momentum distributions for heavy and light species for three points along the purple dotted line shown 
in Fig.~\ref{Fig:DensityDensity}. Left panel: 
The momentum distribution at $\rho=1.19$ 
and $P=-0.06$ (open black square in Fig.~\ref{Fig:DensityDensity}). For both {\it b} (bottom) and {\it a} (top) particles, the distributions have 
a peak at $\vec k = (0,0)$ which corresponds to superfluid behavior.  Middle panel: The momentum 
distribution for the ferromagnetic phase at $\rho=1.33$ and $P=0.08$ (open black circle in Fig.~\ref{Fig:DensityDensity}). For the 
{\it b} particles (bottom), the distribution has a peak at $\vec k = (0,0)$ which corresponds
to  superfluid behavior. For {\it a} particles (top), the distribution 
is uniform corresponding to Mott behavior. Right panel: The momentum distribution at 
$\rho=1.48$ and $P=0.23$ (open black triangle in Fig.~\ref{Fig:DensityDensity}). For both {\it b} (bottom) and {\it a} (top) particles, 
the distributions have a peak at $\vec k = (0,0)$ corresponding to superfluid behavior. 
    }
\label{Fig:Momentum}
\end{figure}

\section{Momentum distribution}\label{Sec:Momentum}
A related experimentally accessible quantity that can distinguish different phases of 
bosons is the momentum distribution. It is defined as
\begin{eqnarray}
  \label{Eq:Momentum} \mathcal \displaystyle N(\vec k) &=&\frac{1}{L^2}\sum\limits_{k,l}e^{i\vec k \cdot (\vec r_k-\vec r_l)}\langle a_{k}^\dagger a_{l}^{\phantom\dagger}\rangle, 
\end{eqnarray} 
with the momentum $k_{x,y}=\frac{2\pi}{L}m$, $m=0, 1, ..., L-1$. The superfluid ground state 
is characterized by a peak at zero momentum, $\vec k= (0,0)$, 
while the Mott insulator phase has an uniform momentum distribution.~\cite{Greiner, Catani} 
Fig.~\ref{Fig:Momentum} displays the momentum distribution of heavy and light particles 
for the ferromagnetic and superfluid phases along the purple dotted line in Fig.~\ref{Fig:DensityDensity}. 
The momentum distribution at zero wavevector, $\vec k = (0,0)$, and $\beta t=50$ as a function of 
polarization for heavy and light species is shown in the top panel. The momentum distribution 
of the heavy particles at zero momentum is small in the ferromagnetic region but large in the 
superfluid phase. The light particles show significantly less variation with polarization
and have a value which is consistently large compared to the ferromagnetic state of the heavy particles,
indicative of a superfluid state. The momentum distributions for heavy and light species for 
three points along the purple dotted line are shown in the bottom panels. 
The left panels  show that for $\rho=1.19$ and $P=-0.06$, heavy and light distributions display peaks 
at $\vec k = (0,0)$ indicating superfluidity of both species. The same happens at
the right panels for $\rho=1.48$ and
$P=0.23$. The middle panel at $\rho=1.33$ and $P=0.08$ 
displays a $\vec k=(0,0)$ peak 
in the momentum distribution of the light species while the momentum distribution 
of heavy particles is uniform. This behavior is consistent with the phase-separated ferromagnetic phase
where the heavy species {\it a} becoming Mott while the light one {\it b} displays regions with 
Mott insulating and superfluid behaviors.

\begin{figure}[!htbp] 
 \centerline{\includegraphics[width=0.5\textwidth]{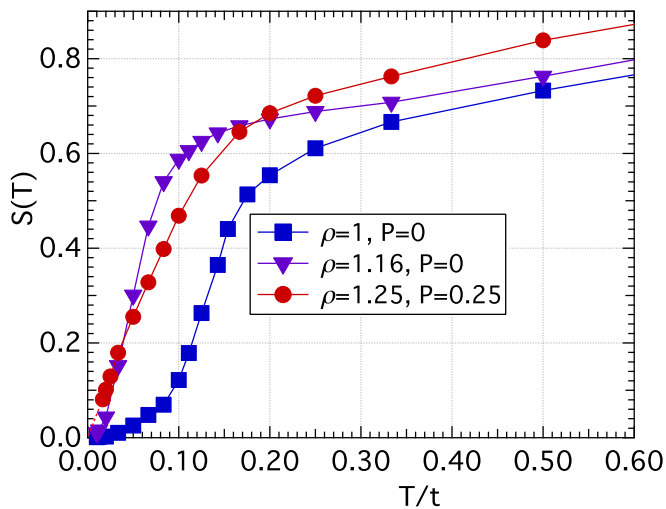}}
  \caption
    {(Color online)
Entropy, $S(T)$, for $L=10$, $t_a=0.08$, $t_b=t=1$, and $U_{ab}=6$, 
as a function of temperature, $T/t$, for three different combinations of  
total densities, $\rho$, and polarizations, $P$. The red circles show the entropy 
of the phase-separated ferromagnetic phase at $\rho=1.25$ and $P=0.25$. The blue 
squares display the entropy of the antiferromagnetic phase at $\rho=1$ and $P=0$. 
The purple triangles show the entropy of the superfluid phase at $\rho=1.16$ for the 
non-polarized system. The data points are based on simulation results, the lines 
are guides to the eye.
    }
\label{Fig:Entropy}
\end{figure}

\section{Entropy} \label{Sec:Entropy}
Reaching the low entropies and temperatures required to observe magnetically 
ordered or Mott insulating phases is still experimentally challenging. 
In the ferromagnetic phase separated region, the superfluid ordering of 
light {\it b} particles can carry most of the entropy,
leaving the entropy of the heavy species in this phase essentially zero. 
Thus the ferromagnetic phase-separated phase can have large entropy. Fig.~\ref{Fig:Entropy} 
shows the entropy for an $L=10$ system calculated following Ref.~\onlinecite{Werner} 
for three different densities and polarizations, $\rho=1$ and $P=0$ (antiferromagnet), 
$\rho=1.16$ and $P=0$ (superfluid), and $\rho=1.25$ and $P=0.25$ (ferromagnet). The entropy 
of the ferromagnetic phase is greater than that of the antiferromagnetic 
phase and similar to that of the superfluid state, especially for low temperatures, 
indicating that it may be more accessible experimentally. The entropy, $S(T)$ is 
calculated by integrating the internal energy per site, $E(T)$, as:
\begin{eqnarray}
  \label{Eq:Entropy} \mathcal S(\beta, n) &=& S(0, n)+\beta E(\beta, n)-\int_0^{\beta} E(\beta^\prime,n)d\beta^\prime,
\end{eqnarray}
where $S(0, n)$ depends on the possible per site occupation of {\it a} and {\it b} 
particles. 

\section{Conclusion} \label{Sec:Conclusion}
By introducing a population imbalance between the two species, we find 
an extended region of phase-separated ferromagnetism in the two-dimensional 
two-species hard-core bosonic Hubbard model. The average local densities show that 
the heavy species has Mott-insulating behavior while the light species 
is phase separated into both Mott insulating and superfluid regions. 
This phase exists for a broad range of temperatures and polarizations. 
In this polarized model we find the optimal superfluid line, 
$\frac{N_a}{2}+N_b=L^2$, where the system shows high transition
temperatures and fully phase-separated regions at low temperatures
with average local densities $n_a\sim 0$ and $n_b\sim1$ on one of the regions, and $n_a\sim1$, 
$n_b\sim 0.5$ on the other. This line exists because the superfluidity of light 
species with $n_b\sim 0.5$ gains the most energy per particle. 
Further the ferromagnetic phase-separated phase is more stable for positive 
polarizations around $\rho =1.25$ and $P=0.25$. When the system deviates far from 
half-filling of the heavy {\it a} particles, the ferromagnetic phase vanishes since it is 
difficult to stabilize small and large phase-separated regions. 
By using finite-size scaling of ferromagnetic susceptibility ratios, 
we find the correlation length exponent $\nu \approx 1$ which is consistent with a two-dimensional Ising ferromagnet. 
Despite its ferromagnetic order, this phase has relatively high global 
entropy, which suggests that its experimental observation in cold atoms
should be more accessible.

\section{Acknowledgment}
This work is supported by NSF OISE-0952300 (KH, VGR and JM) and DMR-1237565 (KH and JM). 
Additional support was provided by the NSF EPSCoR Cooperative Agreement No. EPS-1003897 with additional support 
from the Louisiana Board of Regents (KMT and MJ).  
This work used the Extreme Science and Engineering Discovery Environment (XSEDE), which is supported by the National 
Science Foundation grant number ACI-1053575, and the high performance computational resources provided by the 
Louisiana Optical Network Initiative (http://www.loni.org).


\begin{thebibliography}{5}

\bibitem{Yunoki} S. Yunoki, J. Hu, A. L. Malvezzi, A. Moreo, N. Furukawa and E. Dagotto,Phys. Rev. Lett. {\bf 80}, 845(1998).
\bibitem{Yunoki2} A. Moreo, S. Yunoki and E. Dagotto, Science 2034, {\bf 283} (1999).
\bibitem{Uehara} M. Uehara, S. Mori, C.~H. Chen, S. -W. Cheong, Nature {\bf 399}, 560 (1999).
\bibitem{Elbio1} E. Dagotto, T. Hotta and A. Moreo, Phys. Rep. {\bf 344}, 1-153 (2001).
\bibitem{Coleman} P. Coleman and A.~J. Schofield, Nature, {\bf 433}, 227 (2005).
\bibitem{Si} Q. Si and F. Steglich, Science {\bf 329}, 1161 (2010).
\bibitem{Elbio} E. Dagotto, Science {\bf 309}, 257 (2005).
\bibitem{Jaksch} D. Jaksch, C. Bruder, J. I. Cirac, C. W. Gardiner, and P. Zoller, Phy. Rev. Lett. {\bf 81}, 3108 (1998).
\bibitem{Hofstetter} W. Hofstetter, J. I. Cirac, P. Zoller, E. Demler, and M. D. Lukin, Phys. Rev. Lett. {\bf 89}, 220407 (2002).
\bibitem{Esslinger} T. Esslinger, Annual Review of Condensed Matter Physics {\bf 1}, 129 (2010). 
\bibitem{Timmermans99} E.~Timmermans, P.~Tommasini, M.~Hussein, A.~Kerman, Phys. Rep. {\bf 315}, 199, (1999).
\bibitem{kohler:1311} T.~K\"{o}hler,  K. G\'oral, and P.~S. Julienne, Rev. Mod. Phys. {\bf 78}, 1311, (2006).
\bibitem{Greiner} M. Greiner, O. Mandel, T. Esslinger, T.~W. H\"ansch, and I. Bloch, Nature (London), {\bf 415}, 39 (2002).
\bibitem{Fisher} M.~P.~A. Fisher, P.~B. Weichman, G. Grinstein, and D. ~S. Fisher, Phys. Rev. B {\bf 40}, 546 (1989).
\bibitem{Batrouni} G.~G. Batrouni and R.~T. Scalettar, Phys. Rev. Lett. {\bf 84}, 1599 (2000).
\bibitem{Modugno} F. Schreck, L. Khaykovich, K.~L. Corwin, G. Ferrari, T. Bourdel, J. Cubizolles, and C. Salomon, Phys. Rev. Lett. {\bf 87}, 080403  (2001).
\bibitem{Albus} A. Albus, F. Illuminati, and J. Eisert, Phys. Rev. A {\bf 68}, 023606 (2003).
\bibitem{Modugno1} G. Modugno, L. Khaykovich, K.~L. Corwin, G. Ferrari, T. Bourdel, J. Cubizolles, and C. Salomon, Science, {\bf 297} (2002). 
\bibitem{Ospelkaus} C. Ospelkaus, S. Ospelkaus, K. Sengstock, and K. Bongs, Phys. Rev. Lett. {\bf 96}, 020401 (2006).
\bibitem{Roati} G. Roati, M. Zaccanti, C. D'Errico, J. Catani, M. Modugno, A. Simoni, M. Inguscio, and G. Modugno, Phys. Rev. Lett. {\bf 99}, 010403 (2007).
\bibitem{Thalhammer} G. Thalhammer, G. Barontini, L. De Sarlo, J. Catani, F. Minardi, and M. Inguscio, Phys. Rev. Lett. {\bf 100}, 210402 (2008).
\bibitem{Papp} S.~B. Papp, J.~M. Pino, and C.~E. Wieman, Phys. Rev. Lett. {\bf 101}, 040402 (2008).
\bibitem{Catani} J. Catani, L.~De Sarlo, G. Barontini, F. Minardi, and M. Inguscio, Phys. Rev. A {\bf 77}, 011603 (2008). 
\bibitem{Taglieber} M. Taglieber, A.-C. Voigt, T. Aoki, T. W. H\"{a}nsch, and K. Dieckmann, Phys. Rev. Lett. {\bf 100}, 010401 (2008).
\bibitem{Taie} S. Taie, Y.~Takasu, S.~Sugawa, R.~Yamazaki, T.~Tsujimoto, R.~Murakami, and Y.~Takahashi, Phys. Rev. Lett. {\bf 105}, 190401 (2010).
\bibitem{Altman} E. Altman,  W. Hofstetter, E. Demler, and M.~D. Lukin, New J. Phys. {\bf 5}, 113 (2003).
\bibitem{Soyler} S.~G. S\"oyler, B. Capogrosso-Sansone, N.~V. Prokof'ev, and B.~V. Svistunov, New J. Phys. {\bf 11}, 073036 (2009).
\bibitem{Soyler2} B. Capogrosso-Sansone, S.~G. S\"oyler, N.~V. Prokof'ev, and B.~V. Svistunov, Phys. Rev. A {\bf 81}, 053622 (2010).
\bibitem{Stephen} S. Powell, Phys. Rev. A {\bf 79}, 053614 (2009).
\bibitem{Andrii} A. Sotnikov, D.~Cocks, and W.~Hofstetter, Phys. Rev. Lett. {\bf 109}, 065301 (2012).
\bibitem{Kuno} Y. Kuno, K. Suzuki, and I. Ichinose, J. Phys. Soc. Jpn {\bf 82}, 12450 (2013).
\bibitem{Trousselet} F. Trousselet, P. Rueda-Fonseca, and A. Ralko, Phys. Rev. B {\bf 89} 085104 (2014).
\bibitem{Lv} J. P. Lv, Q. H. Chen, and Y. Deng, Phys. Rev. A {\bf 89}, 013628 (2014).
\bibitem{Kalani} K. Hettiarachchilage, V.~G. Rousseau, K.-M. Tam, M. Jarrell, and J. Moreno, Phys. Rev. B {\bf 88}, 161101(R) (2013).
\bibitem{Monroe} C. Monroe, D.~M. Meekhof, B.~E. King, S.~R. Jefferts, W.~M. Itano, D.~J. Wineland, and  P.~Gould, Phys. Rev. Lett. {\bf 75}, 4011 (1995). 
\bibitem{Popp} M. Popp, J.-J. Garcia-Ripoll, K. G. Vollbrecht, and J. I. Cirac , Phys. Rev. A {\bf 74}, 013622 (2006). 
\bibitem{Li} X. Li, T.~A. Corcovilos, Y. Wang, and D. S. Weiss, Phys. Rev. Lett. {\bf 108}, 103001 (2012).
\bibitem{Ho-Zhou} T.-L. Ho and Q. Zhou, Proc. Natl. Acad. Sci. USA {\bf 106}, 6916 (2009). 
\bibitem{Plich} K. Pilch, A. D. Lange, A. Prantner, G. Kerner, F. Ferlaino, H.-C. N\"{a}gerl, and R. Grimm, Phys. Rev. A {\bf 79}, 042718 (2009).
\bibitem{SGF} V.~G. Rousseau, Phys. Rev. E {\bf 77}, 056705 (2008).
\bibitem{DirectedSGF} V.~G. Rousseau, Phys. Rev. E {\bf 78}, 056707 (2008).
\bibitem{SpaceTime} V. G.~Rousseau and D.~Galanakis, arXiv:1209.0946 (2012).
\bibitem{Banos-etal-2012} R. A. Ba\~nos, A. Cruz, L.A. Fernandez, J. M. Gil-Narvion, A. Gordillo-Guerrero, M. Guidetti, D. I\~niguez, A. Maiorano, E. Marinari, V. Martin-Mayor, J. Monforte-Garcia, A. Mu\~noz Sudupe, D. Navarro, G. Parisi, S. Perez-Gaviro, J. J. Ruiz-Lorenzo, S.F. Schifano, B. Seoane, A. Tarancon, P. Tellez, R. Tripiccione, D. Yllanes, Proc. Natl. Acad. Sci. U.S.A. {\bf 109}, 6452 (2012).
\bibitem{Fisher1} M. E. Fisher, Rep. Prog. Phys. {\bf 30}, 615 (1967).
\bibitem{Fisher2} M. E. Fisher, Rev. Mod. Phys. {\bf 46}, 597 (1974).
\bibitem{Fisher3} M. E. Fisher, Rev. Mod. Phys. {\bf 70}, 653 (1998).
\bibitem{Superfluid} V. G. Rousseau, Phys. Rev. B {\bf 90}, 134503 (2014).
\bibitem{Pollock} E.~L. Pollock and D.~M. Ceperley, Phys. Rev. B {\bf 36}, 8343 (1987). 
\bibitem{Nelson} D.~R. Nelson and J.~M. Kosterlitz, Phys. Rev. Lett. {\bf 39}, 1201 (1977).
\bibitem{Boninsegni05} M. Boninsegni and N. Prokof'ev, Phys. Rev. Lett. {\bf 95}, 237204 (2005).
\bibitem{Werner} F.~Werner, O. Parcollet, A. Georges, and S.~R. Hassan, Phys. Rev. Lett. {\bf 95}, 056401 (2005).
\end{thebibliography}
\end{document}